# Interactive Extraction of High-Frequency Aesthetically-Coherent Colormaps

**Santiago V. Lombeyda**[1]
Center for Advanced Computing Research - California Institute of Technology
**Mathieu Desbrun**[2]
University of Southern California

## Abstract

*Color transfer functions (i.e. colormaps) exhibiting a high frequency luminosity component have proven to be useful in the visualization of data where feature detection or iso-contours recognition is essential. Having these colormaps also display a wide range of color and an aesthetically pleasing composition holds the potential to further aid image understanding and analysis. However producing such colormaps in an efficient manner with current colormap creation tools is difficult. We hereby demonstrate an interactive technique for extracting colormaps from artwork and pictures. We show how the rich and careful color design and dynamic luminance range of an existing image can be gracefully captured in a colormap and be utilized effectively in the exploration of complex datasets.*

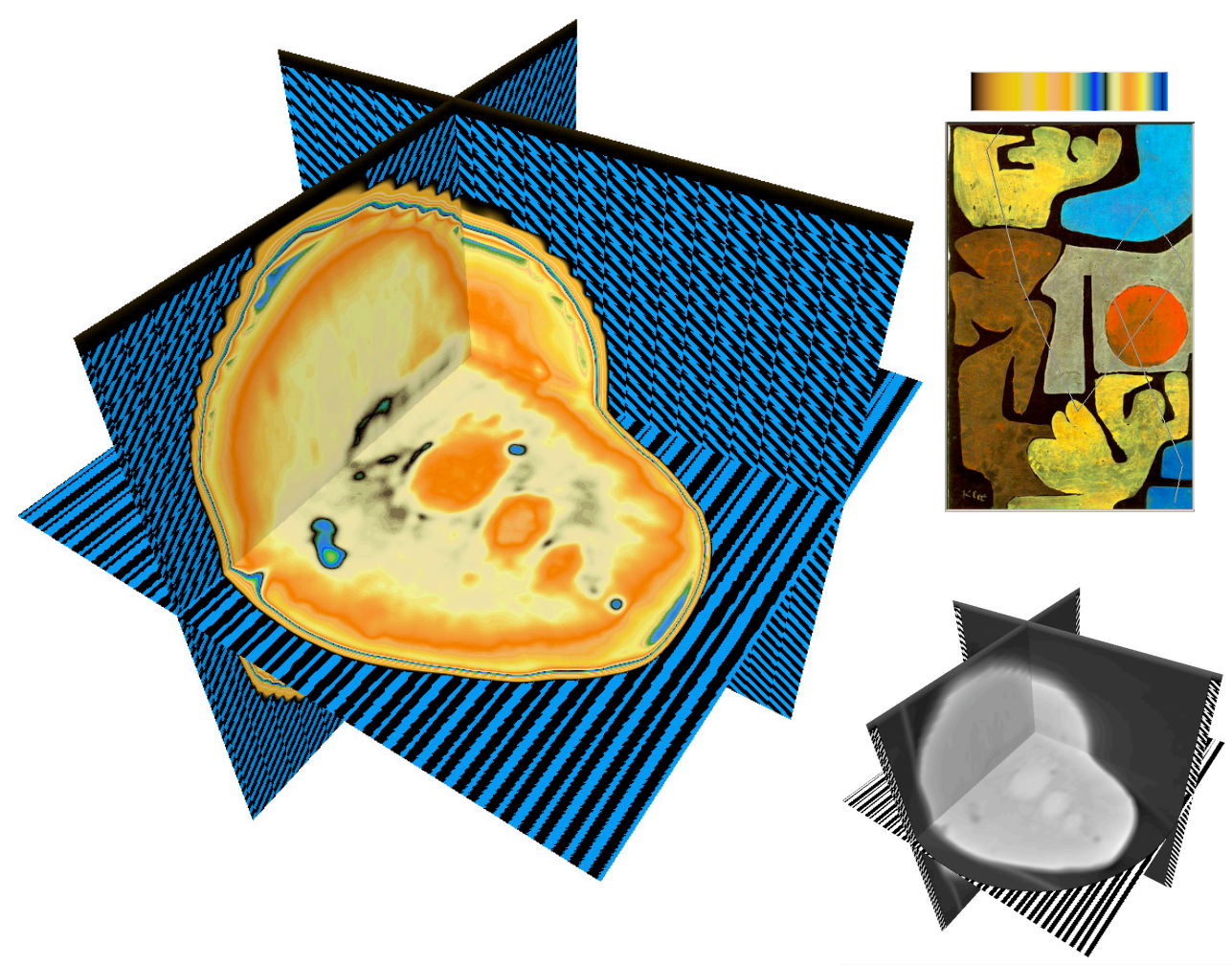

**Figure 1***: Colormap extracted from Klee's "Park of Idols" (1939) versus a standard Grayscale colormap*

## 1 Introduction

Scientific research yields a wide spectrum of large datasets that researchers strive to understand and analyze. Examples of these datasets include results from simulations and scans of fluid flows and shock propagation, magnetic fields, and heat and pressure convection, among many others. Visualization has become the key in understanding these physical phenomena, by visual mining for specific features such as vortices. Visualization plays a similar role for data resulting from high resolution MRI scans or CT scans, used as diagnostic tools in the field of medicine as well as for quantitative studies in paleontology and fossil discovery, where the results are closely studied by researchers looking for specific visual patterns denoting embedded physical structures.

These datasets are commonly studied and explored through the use of slices, volume renderings, isosurfaces, vector displays, and particle traces. In all of these cases, especially with the former two, this is done by mapping the data values to a grayscale intensity or simple hue-spectrum color transfer function. Detecting subtle variances in the data from this straightforward approach is usually quite challenging visually. Furthermore, significant features may be contained within these subtle variances.

The use of an adequate correct color transfer function then becomes key to accelerating the detection of these details, and making them perceivable to the typical discerning end user. This is particularly true in the case of shockwave propagation animations. Here, researchers not only need to distinguish among moving isocontours representing the shockwave fronts, but also need to distinguish one from the next. Being able to follow the movement of these shock fronts is especially interesting when shockwave refraction occurs.

### 1.1 Previous Research

From studies in color perception we know that humans perceive changes in luminance much clearer than changes in hue [Rogo]. Variance in luminance is naturally a more appropriate mechanism for color-coding high frequency information, suggesting that a colormap that offers a high frequency of changes in luminosity is also more likely to yield more information in a visualization. Work such as [Ki02] further supports these premises, and even offers easy ways for a user to calibrate luminance using face-based tools.

In addition to using luminosity variance, we can also use hue and saturation variations to increase the amount of information displayed. Even though research on selectively matching color transfer functions to different types of visualizations has been done, little has been published in this area. This is primarily due to the case by case nature of colormap selection in visualization as well as the difficulty of quantifying an attribute such as colormap efficacy. Nevertheless, as human vision has been found to be extremely adept at recognizing features in images [Will], it is fair to explore how we can optimize this inherent capability by offering rich images that will further aid in the analysis of data.

#### 1.1.1 Scientific Visualization

Much of current research supports correct color choices rather than specific colormaps. Examples of such studies applied to visualization of different areas of science can be seen in

[1] 1200 E. California Blvd., MC 158-79, Pasadena, CA, 91125, SLombey@caltech.edu

[Rogo], [Spar], or [Dere]. There have also been attempts to mathematically aid in the color selection, such that the resulting colors prove distinguishable to human perception [Heal]. However, the user is still thrust with the burden of knowing what color scheme to use and what frequency to place the desired colors in. Thus the time it takes to create a meaningful colormap which proves useful is significantly increased by adding these proven perception constraints.

One solution proposed has been to offer the user a set of predetermined colormaps or carefully steer the user in the creation of one [Berg]. However, prepackaged colormaps and pre-established creation procedures still bring static constraints and limitations. In the worst case scenario, the scientist has to alter the catalogued colormap to the point where the effort spent is comparable to the effort it would take to create a colormap from scratch. Other solutions, suggest semi-automatically generating colormaps though analysis of the data's intrinsic characteristics [Kind]. This, while highly applicable to volume rendering, does not address important issues in color choices, distribution, and perception effects. In fact, several different approaches to generating colormaps were studied and compared as part of a panel [Pfei]. The conclusions from this comparison were highly dependent on the dataset in question, and in most cases best benefited from the knowledge and intuition of the user.

In the end, visualization is the key to reaching a desired audience and conveying the value of research being done. The scientific community understands the value of the aesthetic quality of their results, particularly in differentiating similar research being done at different locations. From this perspective, prepackaged colormaps might fail in offering the user the flexibility to create unique results, and the only other option of creating aesthetically coherent colormaps from scratch requires know-how and time unavailable to most researchers.

### 1.1.2 Turning to the Masters

Looking for a more rigorous, rich, and extensive choice for color schemes we turn to art. [Laid], [Gooc] (see also [Rhyn]), turned their attention towards art to find innovative techniques for visualization. These focus on the knowledge, skills, and specific methods used by artists throughout their art. Similarly, [Inte] introduces the idea of using natural textures extracted from images and applied to scientific visualization in order to allow the user to better relate and understand the data being shown. We, on the other hand, will be taking a closer look at the specific color and composition choices used by artists, and looking to benefit from their choices more directly.

[Corr] characterizes art as "an example [of an image] where the message is contained more in the high-level color qualities and spatial arrangements than in the physical properties of the color." This is especially true in artists such as Claude Monet, where the visual information is hidden in the pure luminance channel, buried underneath the whimsical but precise color use of the artist, as can be seen in Figure 2.

In art we are able to find deliberate color choices and arrangements, conveying information that in most cases transcends the color itself (unless the intentional choice is made otherwise). Thus, our proposal is to use as inspiration these carefully designed color schemes developed and deployed methodically by artists, and use them as our source for our color transfer function generation. Furthermore, these color schemes, as mastered by artists, are endowed intrinsically with the same precepts that have been partially recreated through computational methods, such as in [Heal]. Millennia of the development and refinement of human art assures us that the art which we recognize to be masterful in fact contains a vast "knowledge-base" of color manipulation knowledge and understanding of human color perception.

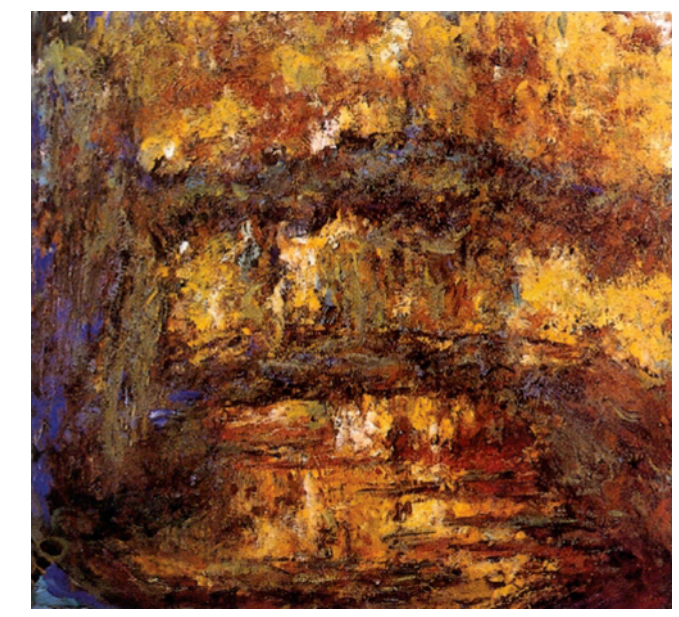
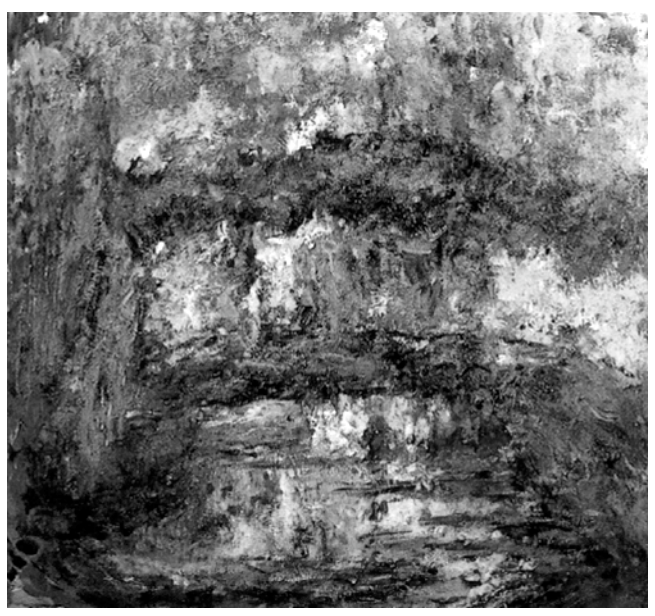

**Figure 2**: *Claude Monet "Le bassin aux Nymphéas", 1923. a) Original painting and b) Luminosity[Dura]*

Whether we are looking at a piece of art or a cover of a magazine, our attention is drawn out by focus point as created by the image composition and color schemes. In fact, perception studies tell us that our brain pieces images together through a series of visual glances at different visual focus points. Color distribution in a picture clearly affects the desired focal points for the image. Work done by [Neig] attempts to automatically find such points of focal interest, and attempt to estimate the paths of vision through focus points by a viewer.

Indeed, artists have been aware of these attention focus paths for quite a long time, and use them to guide an audience through their expression and meaning in their work. A great example of this is the body of work of Paul Klee. Starting with his "Contributions to a Theory of Pictorial Form", Klee attempts to find a relationship between music and painting as a "Rhythmic Exercise" [Ferr]. Thus, highly aware of his composition, art erudites believe Klee planned for deliberate paths of visual focus on his work. This in many ways is the art of composition, essential to all art work. However, in Klee's art pieces, given his writings and body of work, it is easier to infer the specific perception (psychometric) motions he intended for his audience.

Composition along with mastery of the color palette used, as well as careful planning of the different color distributions are clearly beyond the skills and capabilities of the average scientist and visualization designer. Art analysts in fact know that color schemes can be so complex, that just reproducing a color palette as used by a master is an impossible undertaking. In fact it is a primary obstacle in reproducing a masterpiece.

Our efforts center around the ability to used scanned work of art to extract the carefully conceived color schemes, therefore giving the visualization designer a superb "knowledge-base" system encoded in the art work from which to elaborate a unique color transfer function.

### 1.2 Contribution

Our proposition and contribution lies in first accepting that there exist masters in color scheme creation and manipulation. These masters can be found as artists who carefully conceive and implement color manipulation throughout their body of work; whether in a painting, or a photograph. Second, we propose to exploit this erudition by extracting this encoded, carefully selected information, and applying it to scientific visualization. As a result, the end user is able to extract interactively a color transfer function from existing art work that is aesthetically pleasing, highly coordinated, appropriate to the specific task, while displaying a high frequency in intensity and a desired wide spectrum of color as needed.

Our goal is to empower users who want complete creative control over the colormap creation, by offering them a tool which makes the process of achieving a complex, color-rich, and aesthetically coherent colormap quick and easy. Our solution is to allow the user to manually extract colormaps from any desired image.

We will show here the results of the implementation of this tool with a set of art-based images. We characterize art images as photographs or paintings, taken from nature or created by an artist, with intentional choices made for the color distribution and composition. We further suggest, following the work on semi-automatically finding visual focus points [Miau], that the extraction of these colormaps can be automated using visual focus points to extract colormaps from a given art image in a wide range of color variation and luminance frequencies, all with an underlying aesthetically coherent color scheme.

The results will be aesthetically coherent colormaps, where the luminosity demonstrates a high frequency akin to the luminosity visual cues presented in the art pieces (Figure A.1).

We introduce the use of a software tool that extracts colormaps or color transfer functions from images as an alternative to arduous manual creation of color-rich high-frequency colormaps. We use this tool upon images of human produced art so as to benefit from a very deliberate and proven effective color scheme. We then apply the resulting colormaps to scientific data, creating highly detailed and lush visualizations, such as can be seen in Figure 1. The complexity and richness of the resulting transfer functions becomes evident when compared to its grayscale counterpart, especially given that the generation and application of the new colormap took but seconds.

## 2 Color Schemes and Transfer Functions

Figure 3a was extracted from a sequence of MRI scans of a fossil rock collected from the Lincoln Creek formation at exposures in southwestern Washington state. Shown are three orthogonal slices through the data set. The volume reveals what is currently perceived to be a gastropod from the Early Miocene in age (approximately 21-23 Million years old), known as Ancistrolepis Jimgoederti (courtesy of James Hagadorn). Using a linear conversion from the scanned value to grayscale, we are scarcely able to perceive the elliptical features in the bottom cutting plane. These represent the body of the coiled gastropod. This linear mapping corresponds to the colormap seen in Figure 3b. From previous research we know that adding more intensity fluctuation can increase the amount of detail than can be perceived. We do that by increasing the number of grayscale cycles in our colormap to create Figure 3c. Now we can easily observe the intersection of the gastropod with the cutting plane becomes clearer, as three darker ellipses.

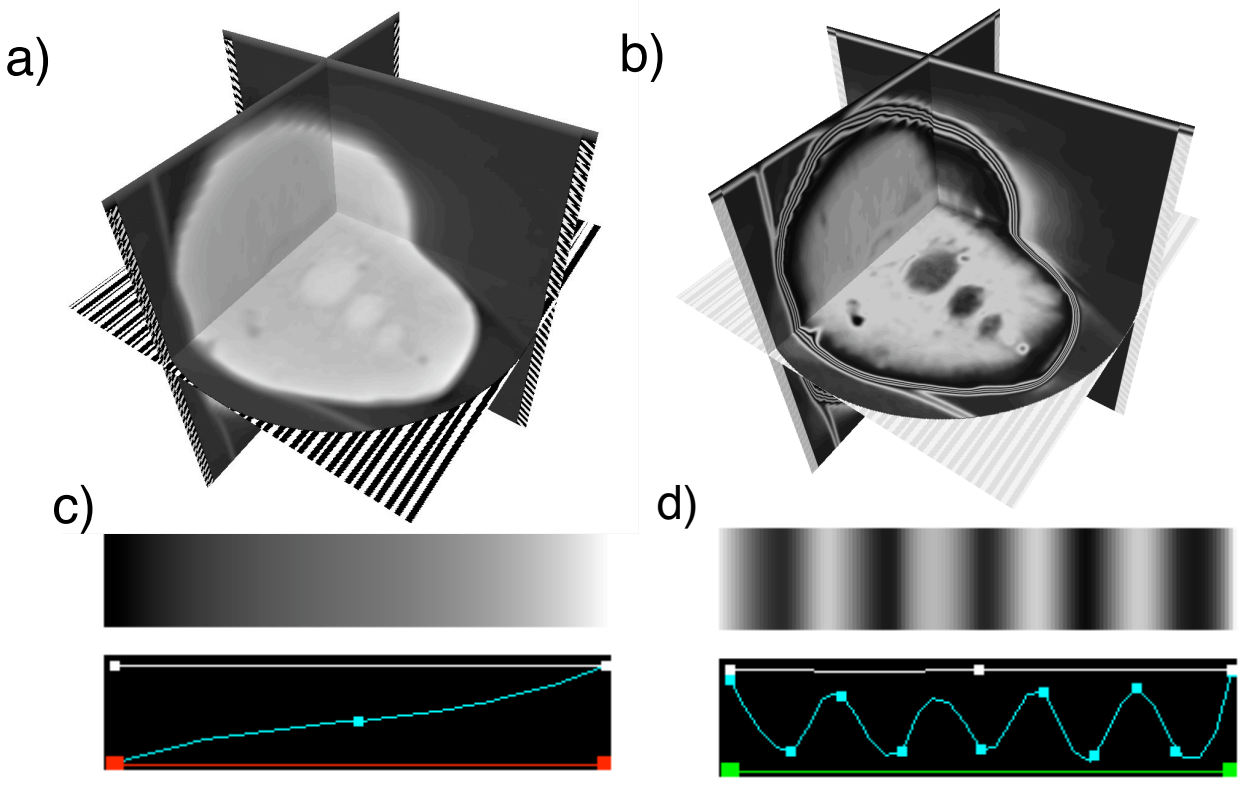

**Figure 3**: *a,c) Orthogonal Cutting Planes through Volume from MRI Scans of gastropod form Early Miocene; b,d) Corresponding Colormaps*

So how many more features can we infer from this image? Is it possible to tell the value at which these features occur? These are obviously very hard propositions to carry out from the images resulting from our initial choices in color transfer functions.

To put it into perspective, it took us approximately seven minutes to perfect the colormap seen in Figure 3d. Much of the time was devoted to making sure the frequency and amplitude of the grayscale cycles appear uniform. The tool used to create the colormap was the Generate Colormap Module under NAG's Iris Explorer [IRIS], a curve-based colormap generation tool. In contrast, a freehand colormap creation tool would have eased the process of creating the cycles. Furthermore, systems such as Adobe Photoshop [Phot] offer manual sketching curve techniques with smoothing options that make this kind of proposition simpler. Nevertheless, once you add separate controls for hue, saturation, and opacity, or red, green, blue, and opacity, the task at hand becomes increasingly difficult.

Basic grayscale intensity ramp colormaps, and simple hue spectrum colormaps, can illustrate a simple data distribution. However, we can improve on these by creating color transfer functions with a higher frequency in intensity when needed. In grayscale, the lack of color variance makes one specific cycle value indistinguishable from the next. As an improvement, the user can vary hue values throughout the colormap. But choosing hue values that are distinct, aesthetically pleasing and coordinated at this point in the process is quite clearly time-consuming, and for the less experienced, quite grueling.

Pre-established colormaps, such as those of [Berg] can greatly accelerate this process. But as the user creates the colormap, they must attempt to reflect a thematic that is congruent with the intention of the visualization in order to avoid conveying wrong information. So, if no pre-established suitable colormap is found which satisfies the user's needs, both of thematic and color variation frequency, the user has to revert to creating the colormap from scratch.

## 3 Color Sampling

Our tool was implemented as a standalone Java application. It takes in an image as input, displays it, and then allows the user to interactively select a path on top of the image from which the colormap will be extracted, as shown in Figure 4. The output is the 256 or 1024 record lookup color table. These values are inferred from evenly spaced sampling points across the length of the path. Notice that this could also be done or suggested automatically using the visual focal points extraction tools discussed earlier, as in the online applet found at [BUVA].

Once a sample line is drawn, sample intervals are computed along the path. The color itself can be sampled in one of three modes. First is the nearest pixel mode, where the pixel closest to the sample interval end is chosen. Second, linear average mode, where colors are assigned a linear weight between neighboring interval ends, and the resulting color is computed as an average. Third, is also a weighted average, where a Gaussian curve is used to assign weights. Figure A.2 illustrates the different results from the different modes.

The first mode, nearest pixel, produces colors that are most faithful to the original image, though it is likely that the graphics engine used when applying the colormap will interpolate between samples to create new colors. Similarly, it also runs the risk to skip some colors of interest that lie between interval ends. Conversely, the sample pixel may be a very atypical color which the user may not be expecting. The second and third methods blend colors, so the resulting color scheme may seem smoother, though it also may seem less drastic.

Thus the resulting colormap from the tool is dependent on its drawn location, the sampling rate (which can be interactively changed), and the color sampling mode. However, since it is up to the visualization engine to do the interpolation for colors between the table entries for the colormap, a few new colors may be introduced.

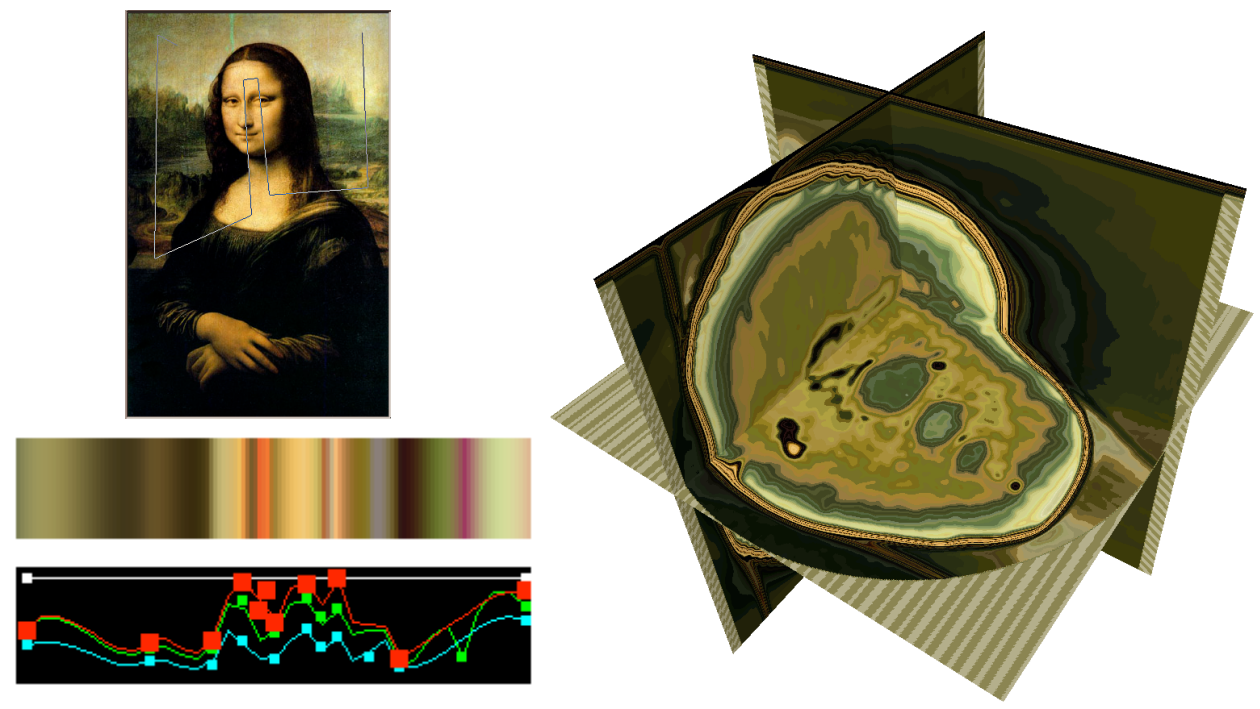

**Figure 5:** *Sample 2: Classic art, color variance demonstrating color clusters.*

## 4 Results

We can characterize the resulting images according to their variation in color content and amount of distinctly present features. Images high in color content will yield colormaps with a high degree of color diversity. Meanwhile, images with large number of clearly demarked features will yield colormaps with high levels of luminance change, as long as the path directly crosses the image feature's boundaries. If the selected path runs a perimeter of one of these image's features, or if the path length is small enough, then the resulting colormap will have low color change deviation. However, it is clear that the user is empowered with all the necessary tools to manipulate the color path in order to achieve any desired result. Notice that if the user decided to create paths spanning a few pixels, then the tool would return a colormap with minimal color/luminosity variation.

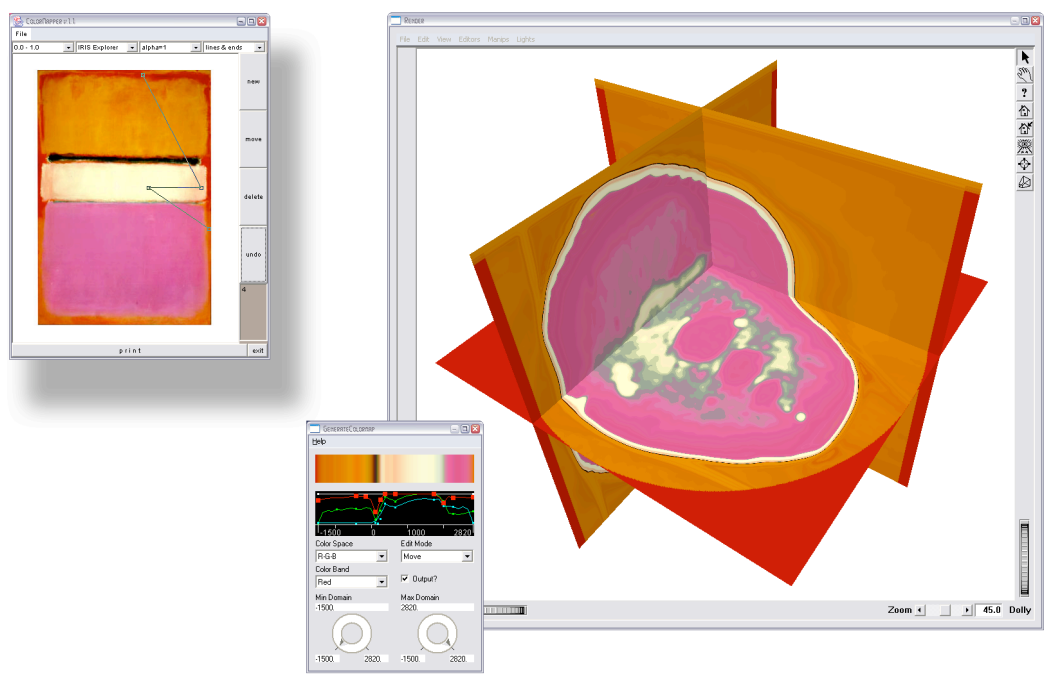

**Figure 4**: *Colormapper tool and Sample 1: Image with simple features, "White Center" by Mark Rothko*

To illustrate some of the different situations encountered, we have chosen a wide range of images to produce colormaps for the gastropod MRI volume shown before.

On the first example, we explore an image mostly devoid of sharp features (Figure 4) as based on a work by Mark Rotko. Two clear regions, apparently low in color and luminance variation, exist. By weaving the path across the color boundary we effectively introduced high color variations into our resulting colormap. Because the main body of the gastropod lies in the higher range values, the ellipses do not appear as clear as possibly wanted. However, the color variation, though slight, is clear enough to mark their location.

The example on Figure 5 illustrates a path set across Leonardo da Vinci's Mona Lisa. The path crosses major visual features, picking up colors from several different color palettes. The resulting image is consistent in color with the source, yet is quite unequivocal in bounding the features. While the whole color transfer function demonstrates continuous variations in luminance, now value regions can be determined. This is expected to be done by the user interactively, with a certain purpose in mind.

Figure 6 samples a piece by El Greco named "The Burial of the Count of Orgaz". The extreme high density of color variation in the different parts of the image means that the path does not need to be long or complicated to gather a colormap high in luminance and color fluctuations. Notice how two different paths, both created extremely quickly, can create radically different results.

Finally, we use an abstract piece by Paul Klee. As Figure 7 depicts, the path cuts diagonally across "The Embrace". Though the color saturation outside the rough thicker black strokes seems bleak, the colormap surprisingly picks up the subtle color variations. Yet again, the natural use of color and dark strokes is clearly conveyed in the resulting visualization. Notice the choice of where the path starts and ends is quite deliberate. Our instinct, which is clearly what the artist intended, was to first focus on the two eye-like features in the mid-top section. By choosing to go from one attention focus point to another, we are sure to encounter large color variations, as color variations and arrangements are one of the criteria that attract our perception and originates focus points. Through the use of this colormap, our resulting image does not only display the equivalent variation in color and luminance, but has created interesting focus points of our own.

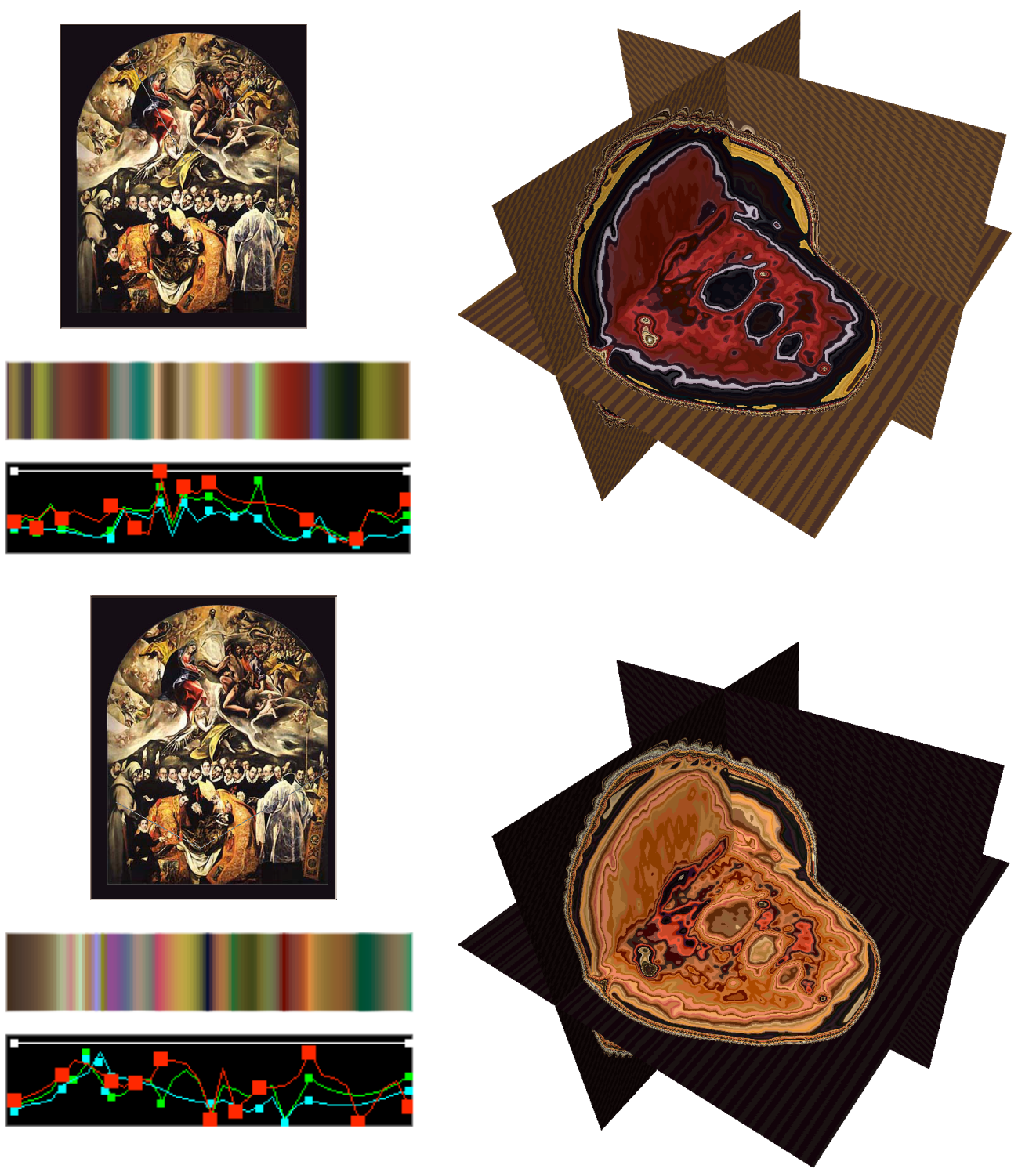

**Figure 6:** *Sample 3: Two different paths using "The Burial of Count Orgaz" by El Greco*

For each of all the presented examples the color map creation took less than a minute. No alterations were needed.

## 5 Future Work

Work on the tool itself includes the added ability to alter the sampling rate density by interactively pulling or pushing tick marks across the path. We have also tested the use of cubic curves instead of straight lines. As expected, the results were not particularly different from those achieved using straight line segments. It actually diminished the ease of the current interface. Color transfer functions obtained from this tool may also be applied to the generation of textures. It is possible to detect the major changes in luminosity from the acquired colormaps, and map these cycles to functionally generated repeating texture patterns (ex. [Soler],[Tong]) to create unique model texturing. The ability to work from a specific image will allow modelers to reproduce natural occurring color schemes on created models without creating an obviously repetitive pattern.

Another interesting area of development will be to allow the tool to suggest paths as resulting from visual paths among attention focus points [Miau][BUVA]. As the system creates a series of complex colormaps in this semi-automatic mode, it would be then possible to create spreadsheets/tables of different applications of unique colormaps for fast data study [Jank].

For different examples using natural photography [Afga] as well as different datasets see additional figures in Appendix B.

## 6 Conclusions

All results demonstrated throughout this paper took under a minute for the sole creation of the colormap. Given the results, we believe that the tool is extremely valuable. It is our belief that such a tool will be useful under most visualization endeavors. It also seems clear from looking at the results that some intrinsic artistic/emotional values of the original images are carried over onto the applications. Further work, evaluating color patterns and color density values between the original image and results from diverse paths, may prove interesting and valuable in trying to mimic an artist's ability to portray specific scenes.

In addition, for sciences where actual images captured from experiments or actual structure exist, this type of colormap extraction may serve as a quick low overhead (versus texture maps) method to create valuable color-coding of the data segments. This may include medical data (ex. picture of a tumor amidst healthy tissue), paleontology data (ex. illustration of a fossil's estimated epidermis), or fluid dynamics simulations (ex. pictures from field experiments with gases, combustion, jets).

It is not possible to generalize and comprehensively measure the effectiveness of colormaps across all implementations of scientific visualization, for the obvious reason that individual visualizations carry different goals and needs. However, it is in fact this case-by-case nature of visualization that has created the need for flexibility in colormap creation. We therefore perceive our approach has not only met this need to create successful custom color transfer functions, but also offers the ability to create them quickly and with ease.

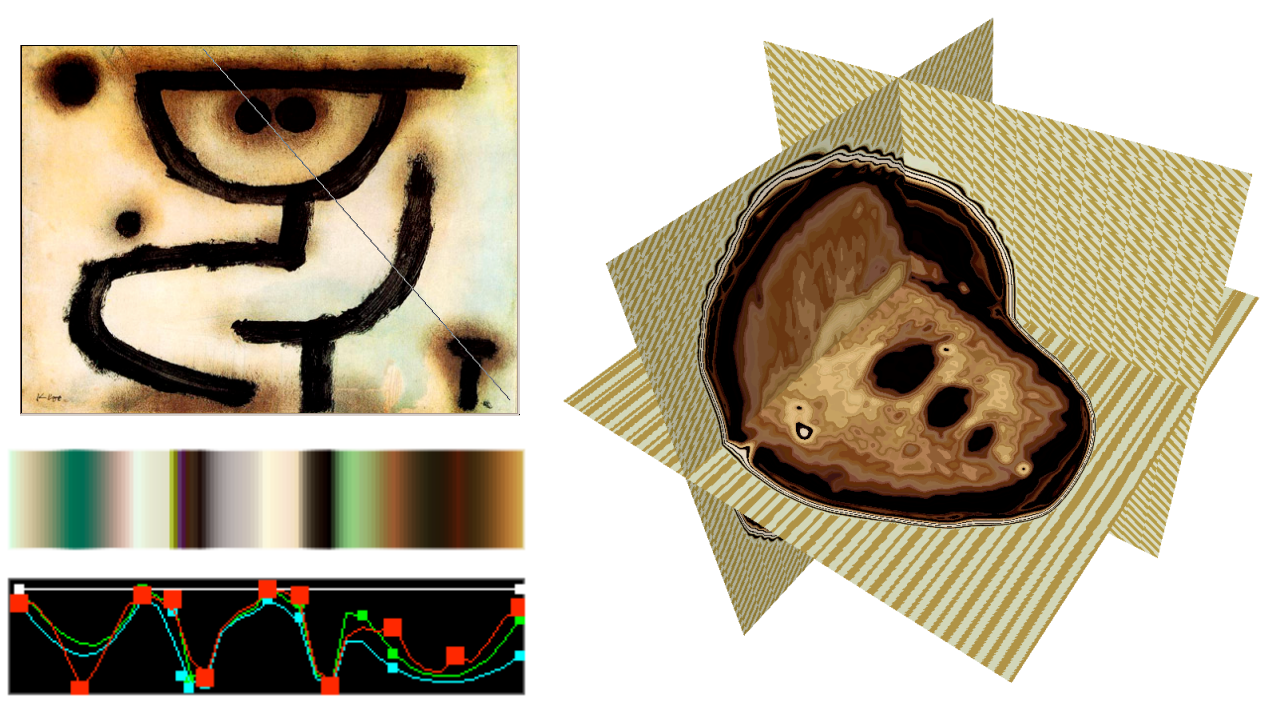

**Figure 7**: *Sample 4: "The Embrace" by Paul Klee - Abstract painting*

## Acknowledgments

Our thanks go to Vicent Robins, Art Center School of Design in Pasadena, for his contributions on Art History and his detailed insights into Color Theory and Paul Klee, and Fredo Durand for his insight and aid. We would like to thank James Hagadorn, from Division of Geological & Planetary Science at the California

Institute of Technology for giving us access to MRI scans of the fossils used in this paper. We would also like to thank Paul Dimotakis (California Institute of Technology) for giving us access to the jet stream scanned data, and with whom we collaborate on their visualization efforts. Finally our thanks to Ashwini Nayak for valuable help editing this paper.

This work is partially funded by the ASCI/ASAP Center of Excellence, the Large Scientific Data Visualization Program (an NSF grant), and Center for Advanced Computing Research support. Photos courtesy of Agfa [Agfa].

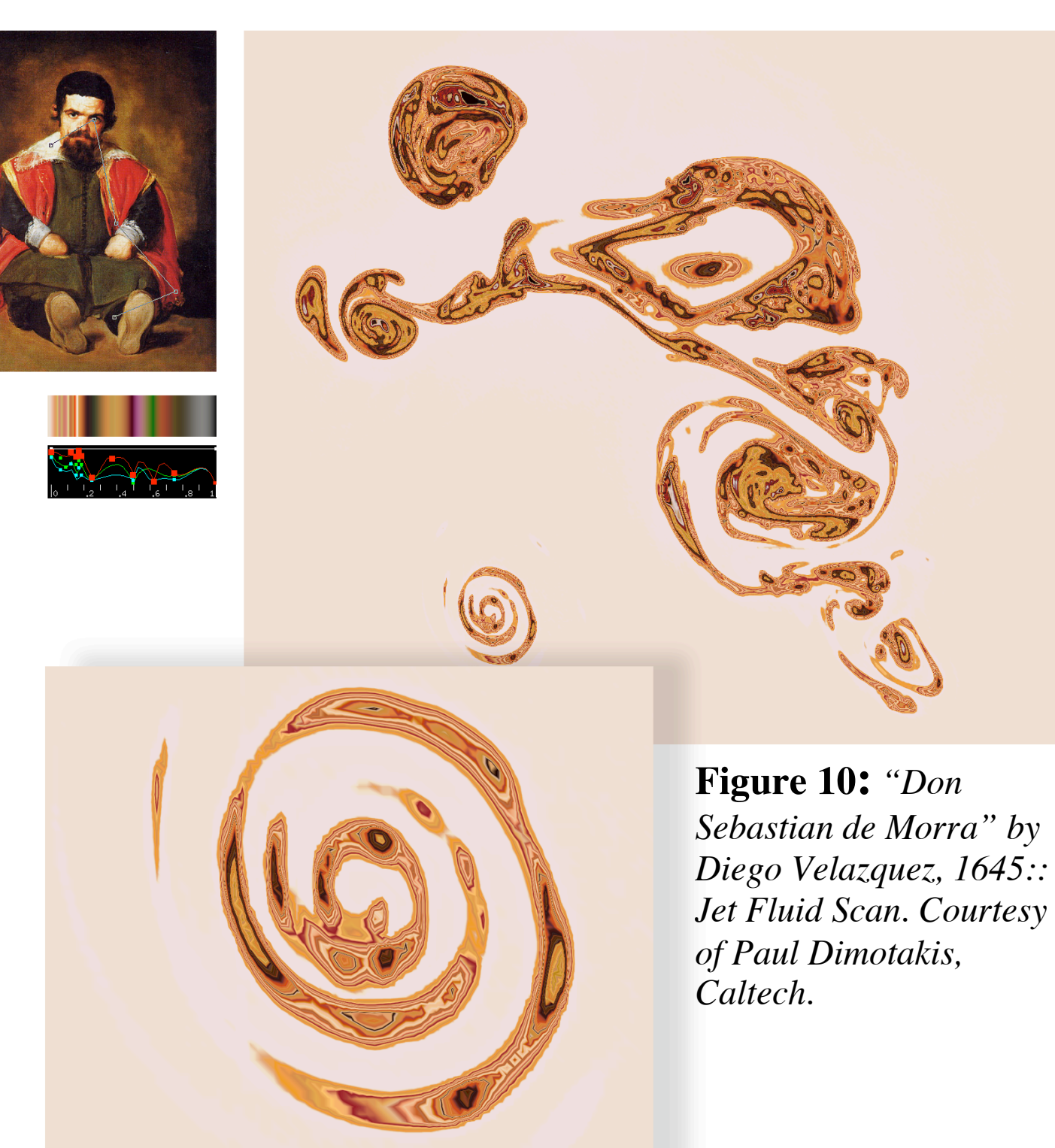

**Figure 10:** *"Don Sebastian de Morra" by Diego Velazquez, 1645:: Jet Fluid Scan. Courtesy of Paul Dimotakis, Caltech.*

## Appendix A - Illustrations

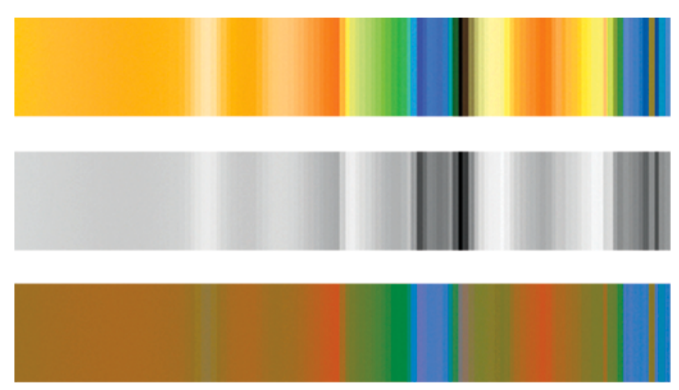

**Figure A.1**: *a) Colormap extracted from Klee's "Park of Idols" b) luminocity component c) color component.*

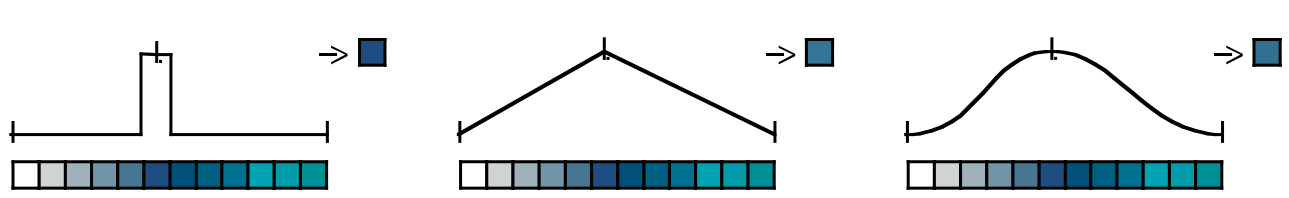

**Figure A.2**: *a) closest pixel color sampling, b) linear sampling, c) Gaussian curve based sampling*

**Figure B.1-2:** *Further Photographic Examples. [Afga]- Notice the first image has little hue variance, yet its high luminosity frequency creates a very effective colormap.*

## Appendix B - Further Examples

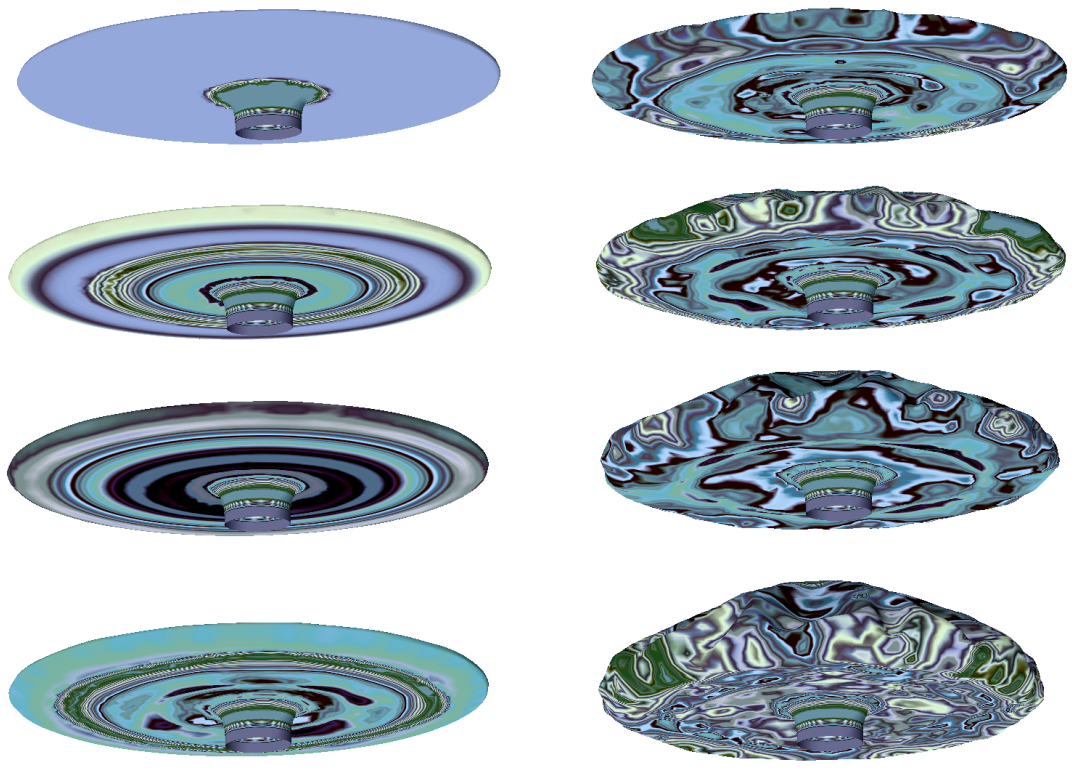

**Figure B.3:** *"La Vie" by Pablo Ruiz Picasso, 1904:: Pressure front propagation in a thin-shell simulation of an airbag- early frames. Courtesy of Fehmi Cirak.*

**Figure B.4:** *"Composition V" by Modrian, 1975:: AMR Mesh Level debugging Visualization. Courtesy of the ASCI/ASAP Center at Caltech.*

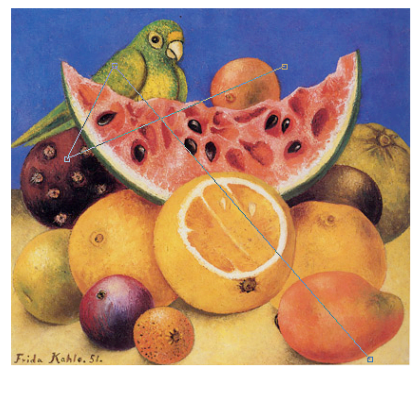

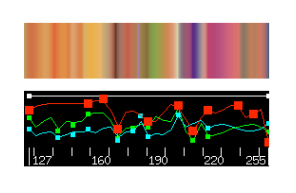

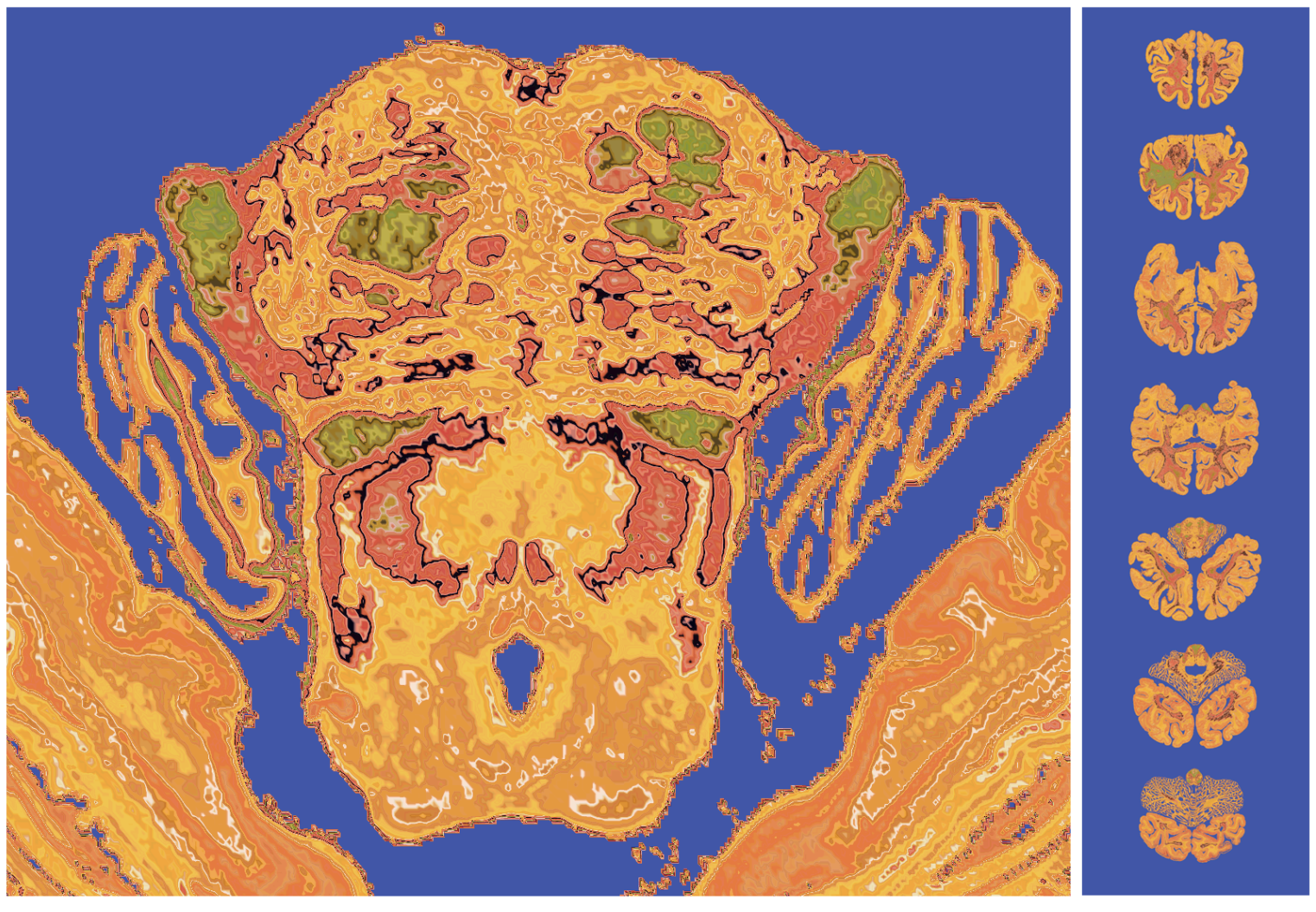

**Figure B.5:** *"Still Life with Parrot" by Frida Kahlo, 1951:: Scans of slices of a monkey's brain. Courtesy of John Allman Lab, Caltech.*